# 2W/nm Peak-power All-Fiber Supercontinuum Source and its Application to the Characterization of Periodically Poled Nonlinear Crystals

B. A. Cumberland, J. C. Travers, R. E. Kennedy, S. V. Popov and J. R. Taylor

Femtosecond Optics Group, Physics Department, Imperial College London, SW7 2BW, United Kingdom (e-mail: b.cumberland@imperial.ac.uk)

**Abstract**

We demonstrate a uniform high spectral brightness and peak power density all-fiber supercontinuum source. The source consists of a nanosecond Ytterbium fiber laser and an optimal length PCF producing a continuum with a peak power density of 2 W/nm and less than 5 dB of spectral variation between 590 to 1500 nm. The Watt level per nm peak power density enables the use of such sources for the characterization of nonlinear materials. Application of the source is demonstrated with the characterization of several periodically poled crystals.

**Introduction**

Photonic crystal fibers (PCF) with high nonlinearities and customizable dispersion profiles [1, 2] have been widely used for supercontinuum generation. Considerable interest has been driven by a diverse range of potential applications including frequency metrology [3], optical coherence tomography [4] and biomedical imaging [5]. Many applications, particularly those involving nonlinear optical processes, require high spectral power densities and single-transverse mode outputs. The highest average spectral power densities of greater than 10 mW/nm have been demonstrated with CW pumping of PCFs with fiber lasers. Enhanced power densities can be obtained with high peak power pumping of PCFs or tapered PCFs [6, 7] over a spectral range as broad as 0.3 to 1.8 µm using picosecond or femtosecond sources. However, the peak power density of such sources may vary considerably across the continuum spectrum not only due to the average spectral power variation (femtosecond and picosecond pumping) but also due to dispersion related variation of the pulse duration across the spectrum. The resulting, often over 10-fold, peak power density changes limit the useful spectral range of such sources in applications where evaluation of linear [8] and especially nonlinear optical responses are involved.

The effect of dispersion on pulse durations is greatly reduced with the use of nanosecond pulse durations. Pumping a PCF near the zero dispersion wavelength with its length adjustment can lead to uniformly high peak power density distribution across the whole continuum spectral range. Comparing to micro-chip or solid-state laser pumping [9] of PCFs that generally result in over 10dB spectral fluctuations and mW/nm scale peak power outputs, the use of all-fiber format pumping with low-loss splicing to the optimized PCF significantly enhances environmental stability of the source and can allow for considerably higher, Watts level, average pump power budgets and Watts/nm output spectral densities.

Here we demonstrate an all-fiber nanosecond supercontinuum source that produces a peak power density of 2W/nm (0.18 mW/nm average) from 590 nm to 1500 nm. The source consists of a kW-peak power single-mode nanosecond Ytterbium pump laser spliced to a PCF whose length is optimized to provide the required spectral and peak power density coverage in the region of interest for the desired application. We apply the high power broadband source to the characterization of the quasi-phase-matching wavelength of periodically poled crystals. The use of such a broadband source offers significant advantages over a tunable laser, as it allows a simple method of single shot characterization over a broad spectral range at Watt-level peak powers. Apart from environmental and spectral power density stability, the full fiber integrated format of the resulting source can make it a robust, low maintenance, low cost diagnostic device.

**Continuum Generation**

The supercontinuum PCF was pumped by a nanosecond pulsed Ytterbium fiber laser (IPG Photonics) in the anomalous dispersion regime. At a pulse repetition rate of 5 kHz the pump laser produced 6.7 kW of peak power with a single transverse mode output ($M^2 < 1.1$) in 18 ns pulses at 1062 nm. The PCF had a zero dispersion wavelength of 1.04 µm, core diameter of 4.9 µm and a nonlinear coefficient of ~13 $(W\ km)^{-1}$. The PCF was directly spliced to the Ytterbium fiber laser enabling a high coupling efficiency of the pump light and suppression of higher order transverse mode excitation. A low loss in-line isolator was employed to protect the laser from possible back reflections from the splice. The typical splice loss, without using any intermediate mode-matching fibers, was 1.5 dB and was obtained by developing an optimized splicing procedure with a filament fusion splicer.

The nanosecond-pumped continuum formation is driven by a combination of nonlinear processes in the PCF. A large number of observed supercontinua, particularly in the femtosecond pump regime, are attributed to soliton fission, dispersive wave generation and cross-correlation effects [9]. However in this case, with pulse durations of 18 ns, the soliton number of the order of $10^6$ is dominated by a very high dispersive length, $L_D$, of around $10^{10}$ m. The corresponding soliton fission length ($L_{fis}=L_D/N$) is of the order of $10^4$ m. Given that our PCF is a few tens of meters, we do not expect that such fission effects will play a large role. In contrast, the effective MI length (~16 $L_{NL}$) is a few 10s of cm – much shorter than our fiber length – and therefore we suggest that modulation instability is the primary mechanism for initiating the continuum. The resulting, spectrally broad solitons, are enhanced towards the longer wavelengths due to the strong Raman gain at the first Stokes wavelength and develop into a continuum via Raman self-scattering.

We believe that four-wave mixing is the dominant effect generating the short wavelength side [9-14].

In order to evaluate the continuum flatness a cut back of the PCF from 62 m to 2.8 m was performed. The contribution of particular nonlinear effects to the supercontinuum generation can be traced from the cutback results shown in Figure 1. For the 3m length of fiber the continuum generation is strongly driven by Raman effects while for the lengths of PCF over 10m, wavelengths shorter than the pump are also generated. The spectral power density in the infrared falls as the PCF length is increased due to the soliton self-frequency-shift driving power transfer to longer wavelengths where the loss of the fiber is higher. For the longest PCF lengths employed, around 60 m, the fiber losses lead to peak power density reductions across the whole spectral range. The dip at 1.38 µm corresponds to the water peak loss.

Depending upon the required application, spectral power densities can be controlled by tailoring the PCF length, dispersion profile [11] and the coupled pump powers. As an example we characterized a number of periodically poled crystal samples, with second harmonic quasi-phase-matching (QPM) wavelengths ranging from 700 to 1550 nm. The PCF length corresponding to the highest most uniform brightness in this spectral range was used. The 17 m long PCF produced a continuum with less than 5 dB peak-to-peak power variation from 590 to 1550 nm and a peak power density of 2.0 W/nm within this range.

**Non-linear Crystal Characterization**

The setup is shown in Figure 2. Periodically poled $KTiOPO_4$ (PPKTP), $LiNbO_3$ (PPLN) and $LiTaO_3$ (PPLT) with poling periods from 2.4 to 24.3 µm (50:50 duty factors for the first order SHG) were assessed and an edge spectral filter was used to facilitate easy identification of the second harmonic signal. Although the pump source was unpolarized, due to the high peak power density, no polarizing elements were needed before the crystals. The crystal samples were 10 mm long which corresponded to QPM bandwidths for second harmonic generation (SHG) of 0.05 to 2.40 nm (FWHM).  A 75mm focusing lens was used producing a 100 µm spot diameter in the crystal.

Typical spectral traces of the second harmonic signals from the tested crystals are shown in Figure 3 for: (a) congruent PPLT, Λ = 2.46 µm, $\lambda$pump = 770 nm; (b) congruent MgO doped PPLN, Λ = 10.23 µm, $\lambda$pump = 1178 nm; (c) PPKTP, Λ = 24.28 µm, $\lambda$pump = 1550 nm; taken at room temperature (21°C). The traces were taken on an Anritsu MS9030A / MS9701B optical spectrum analyzer (OSA).

The second harmonics were readily identifiable for all the crystal samples (766, 1170 and 1534 nm respectively). In the case of the PPKTP (Fig.3c) additional signal was present at 515 nm that we attribute to third harmonic generation of 1545 nm. In such an arrangement, the spectral position of the signals versus crystal temperature data can allow calculation of Sellmeier equation coefficients. In reverse, once the poling period is known, the corresponding Sellmeier coefficients can be deduced for each of the particular ferroelectrics by accounting for the crystal temperature, QPM wavelength and thermal expansion coefficient. A temperature sweep of the crystal pumped by the high brightness continuum source enables verification of the linear thermal expansion coefficient, as demonstrated for the MgO cPPLN crystal (Fig.4).

More detailed information on the poling period, duty factor, poled domains uniformity and Sellmeier coefficients could be obtained by using shorter crystal samples with wider QPM bandwidths. Under these conditions the QPM profile deviation can be more easily monitored.

**Conclusion**

In this paper we demonstrated a compact, robust, efficient, diffraction limited output, all-fiber supercontinuum source capable of producing a peak power density of 2W/nm (0.18 mW/nm average) in the spectral range from 590 nm to 1500 nm. This optimized PCF length source yielded peak-to-peak spectral power variation of less than 5 dB which makes it suitable for non-linear materials characterization in the whole continuum range. The single-spectral shot characterization of nonlinear optical SHG crystals was demonstrated by the use of several samples of periodically poled ferroelectrics.

**Acknowledgements**

This work was supported by the Engineering and Physical Sciences Research Council (EPSRC) under Grant EP/514408. B. A. Cumberland, J. C. Travers and R. E. Kennedy are funded by EPSRC studentships. J. R. Taylor is a Royal Society Wolfson Research Merit Award holder.

**Figures**

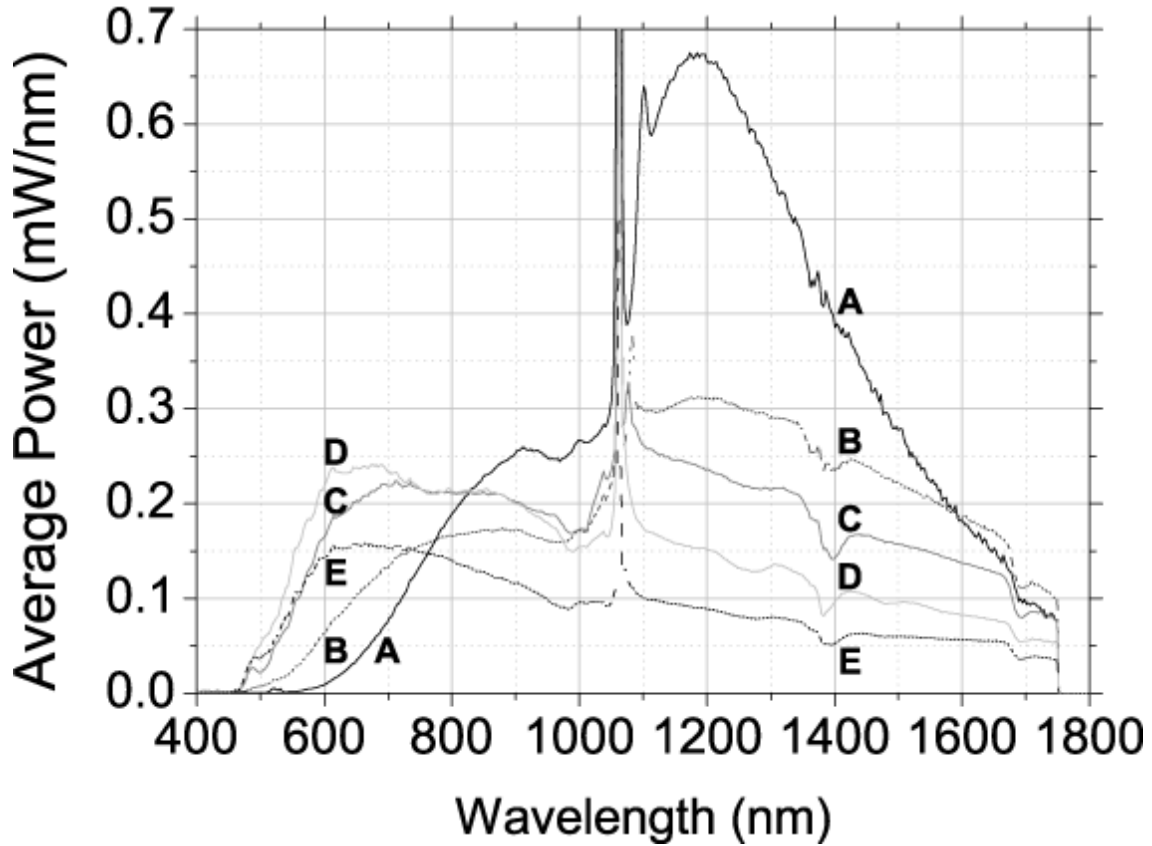

Fig. 1. Supercontinua evolution during a PCF cut-back experiment: (A) 3 m, (B) 10 m, (C) 17 m, (D) 27 m, (E) 62 m. The curves are normalized to output power that does not take into account spectral power outside the OSA coverage.

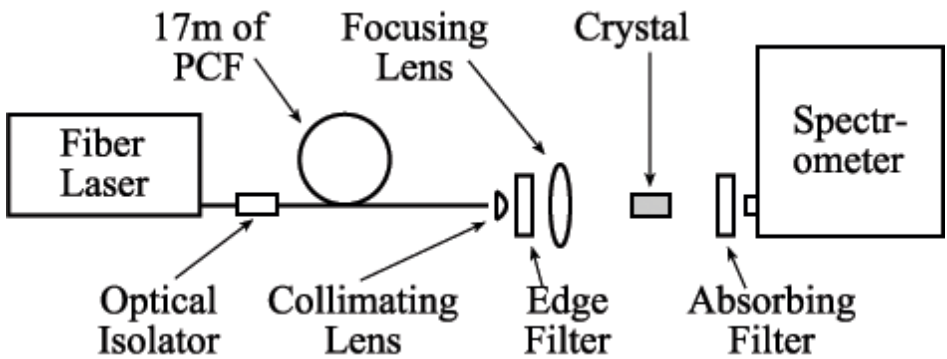

Fig. 2. Experimental setup for nonlinear crystals characterization.

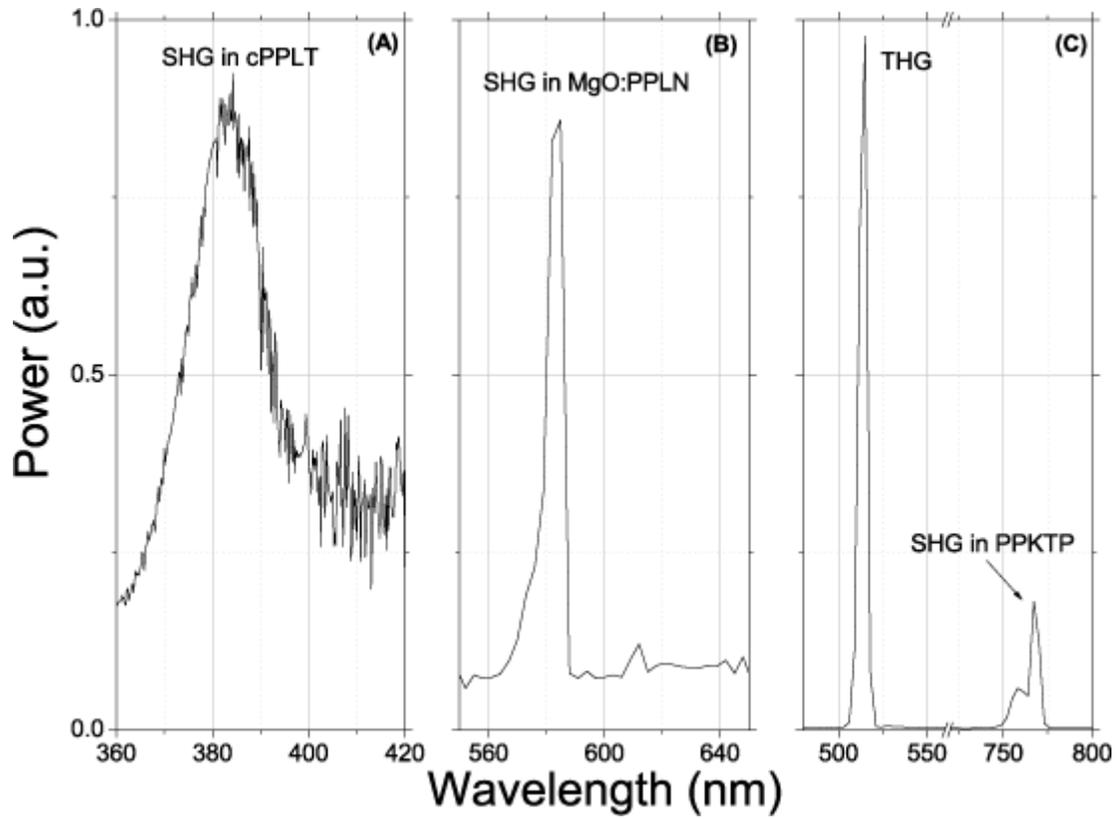

Fig. 3. Spectra of QPM harmonics signals generated in (A) cPPLT, SHG of 766 nm; (B) MgO cPPLN, SHG of 1170 nm (C) PPKTP, SHG of 1534 nm and THG of 1545 nm.

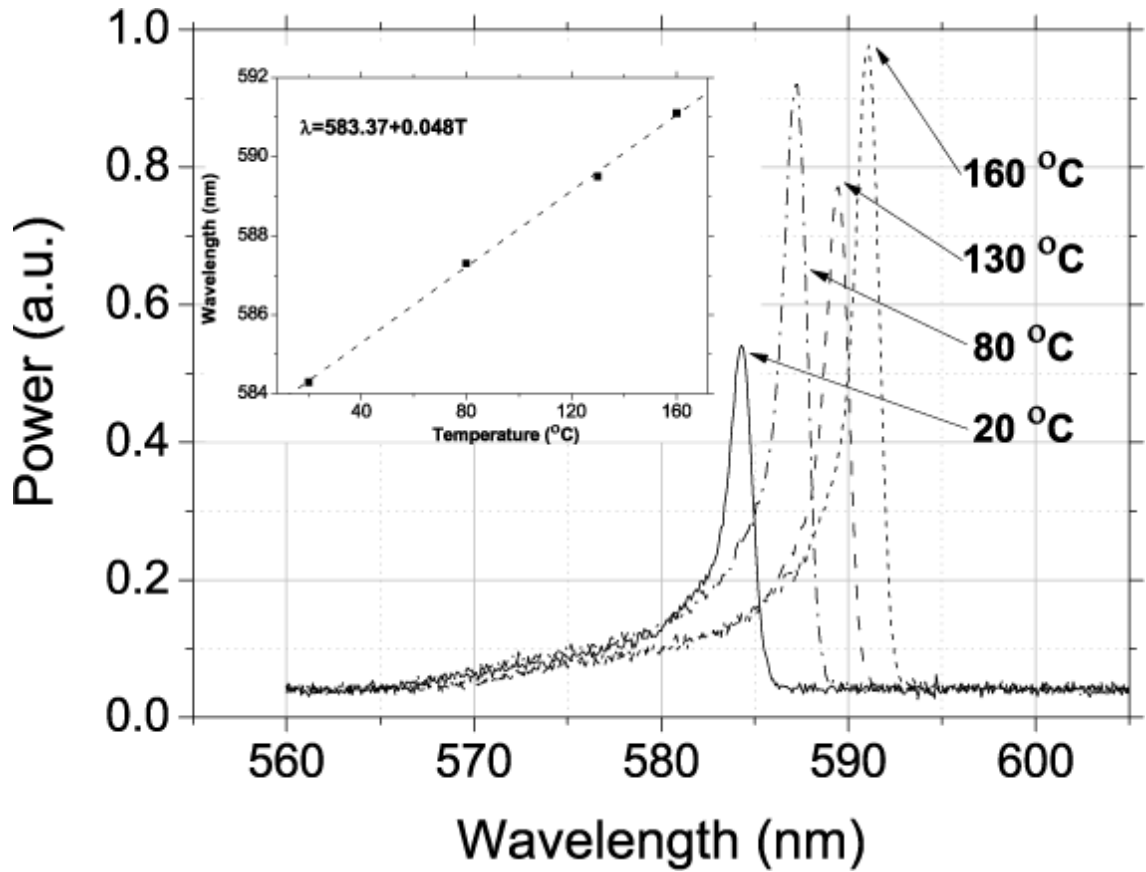

Fig. 4. Dependence of the QPM SHG output spectra of the MgO cPPLN crystal (SHG of 1170nm) on the crystal temperature variation 20 to 100 °C. Inset; QPM wavelength vs temperature.